# Ultra-sensitive Experimental Detection of Chiral Meso-structures by Orbital Angular Momentum of Light

Jincheng Ni[1], Zhongyu Wang[1], Zhaoxin Lao[1], Yanlei Hu[1], Kun Huang[2], Shengyun Ji[1], Jiawen Li[1], Zhixiang Huang[3], Dong Wu[1], Chengwei Qiu[4], and Jiaru Chu[1]

[1]CAS Key Laboratory of Mechanical Behavior and Design of Materials, Department of Precision Machinery and Precision Instrumentation, University of Science and Technology of China, Hefei, Anhui 230027, China.
[2]Department of Optical Engineering, University of Science and Technology of China, Hefei, Anhui 230026, China.
[3]Key Laboratory of Intelligent Computing and Signal Processing (Ministry of Education), Anhui University, Hefei, Anhui 230039, China.
[4]Department of Electrical and Computer Engineering, National University of Singapore, Singapore 117583, Singapore.
Correspondence and requests for materials should be addressed to Y. H. (huyl@ustc.edu.cn) or to K. H. (huangk17@ustc.edu.cn) or to D. W. (dongwu@ustc.edu.cn)

**Circular dichroism (CD) induced by spin angular momentum of light is vital to investigate the chirality of microscopic objects such as molecules, proteins and metamaterials. However, orbital angular momentum (OAM) of light failed to directly implement an interaction with chiral molecules, which remains poorly understood. Even with the enhancement of unique plasmonic nanoparticle aggregates, the helical dichroism (HD) is still at a small level of ~0.6%. Here, we experimentally report the direct observation of a giant HD at the level of ~120% by matching the scales of chiral microstructures and OAM beams. Our results reveal a strong interaction between OAM beams and chiral microstructures in terms of their distinct reflectance, resulting in sign-opposite HD values to**

**distinguish their chirality. Such unique HD phenomenon is investigated experimentally under various diameters and helical pitches of chiral structures, showing the robust performance. This HD technique can widely extend the responding areas of chiral spectroscopy in biology, material science and nanotechnology.**

# Introduction

Like rotating particles, the total angular momentum of massless photon can be divided into spin and orbital angular momentum[1]. More than one hundred years ago, Poynting[2] reasoned that the spin angular momentum (SAM), defined values (**S**=$\pm\hbar$, $\hbar$ is the reduced Planck constant), was related to the left- or right-handed circular polarization. However, only two decades ago, Allen[3] et al. recognized that the orbital angular momentum (OAM) was carried by a paraxial Laguerre-Gaussian beam with helical phase wavefronts. The OAM of light can be evaluated by **L**=$l\hbar$, where the integer $l$ is the topological charge and its sign indicates the handedness of helical wavefront. These orthonormal OAMs with unbounded values have been extensively used in optical tweezers[4], quantum entanglement[5,6], optical communications[7], micro/nanofabrication[8], and sensing[9].

Chirality, describing the symmetry properties of an object, is quite ubiquitous in nature. A chiral object cannot be superposed with its mirror by rotations or translations, such as hands, DNA, proteins, seashells, or spiral staircases[10]. Apart from direct observation of geometric features, three-dimensional contour of chiral structures can also be probed via a different response to right- and left-handed circularly polarized (RCP and LCP) light[10,11]. The chiral optical response in terms of a polarization-dependent extinction coefficient is called as circular dichroism (CD), which is a powerful tool to investigate molecules in biology, chemistry, and material science. Yet most natural chiral molecules generally possess extremely weak CD response, where the dimensions of chiral molecules are much smaller than the wavelength-scale helical pitch of circularly polarized light[12]. Recent works have discussed to enhance the CD response by matching the chiral scales of materials and the helical pitch of circularly polarized light, such as chiral metamaterials[11,13], or superchiral light[14]. Nevertheless, the CD spectroscopy using SAM of light is still inconvenient and expensive for chiral materials with different sizes, which requires tunable laser sources and assorted optics for the scale matching. Therefore, developing a wavelength-independent approach for chiral detection might benefit the fields such as biology, chemistry and optics.

Consequently, the OAM of light is taken as an alternative way to probe the chirality of microscopic objects. The attempt to detect chirality of objects by using the OAM of light has failed in a long period[15,16]. Nevertheless, substantial efforts have recently been made to face the challenge. For example, OAM has been used to induce CD in an achiral nano-aperture with its angular momentum broken the mirror symmetry of input beams[17]. Furthermore, by using unique plasmonic nanoparticle aggregates to enhance the electric quadrupole fields, OAM can engage with chiral molecules to discriminate enantiomers[18]. However, such an indirect approach employing other auxiliary structures leads to the weak helical dichroism (HD~0.6%) and the underlying mechanism of HD remains cursory. Therefore, it is of practical significance to propose a new paradigm that can directly induce a strong HD for the discrimination of chiral structures.

Here, we show for the first time direct observation of a giant HD arising from the strong interaction between chiral microstructures and focused OAM beams. The HD response is symmetric for two enantiomers, where the HD values are equal in magnitude but opposite in sign. We experimentally obtained the maximum HD of 120%, which is enhanced by 200 times compared with previous report. The dependence of HD response on the dimension of chiral structures has been investigated, exhibiting a robust behavior to extend its applications beyond the regimes explored by CD response, which also shows the capability of remotely measuring the chirality of macroscopic objects in astronomy.

## Results and discussion

**Principles of strong HD effects.** The mechanism of our proposed HD is clearly distinguished from the traditional CD spectroscopy that employs the different response between LCP and RCP light over a range of wavelength, as shown in Fig. 1a. Firstly, the peak wavelength of CD spectroscopy is comparable or larger than scales of chiral structures[19,20], whereas the HD locates at matching the scales of chiral structures and OAM beams (see Fig. S3). Secondly, compared to SAM with only two state, the theoretically unlimited OAM provides another dimension for detecting the chiral structures instead of wavelength. Therefore, the HD spectra of different chiral structures can be available by implemented as a function of OAM values to avoid changing the wavelength $\lambda$. Thirdly, the HD spectra are located at the diagonal section of CD response region (Fig. 1b). The peak wavelength of CD spectroscopy is red-shifted to a larger diameters of chiral structure, but keeping in the zone of $D<\lambda$[11,19,21-25]. In this study, the HD spectra of chiral structures are generated at the wavelength smaller than the diameters of chiral microstructures. Namely, the HD peaks are in the zone of $D>\lambda$, which is still an unexplored region. The simple and intuitive diagram can be used to understand the CD and HD response behaviors on dimensions of chiral structures. Notably, all HD measurements in our proposal can be achieved at the same wavelength, which reduces the requirement of broadband sources and assisted waveplates.

To understand the origin of the HD response, we study the interaction between OAM beams and chiral structures. Distinguishing from previous results[16,18], the dimensions of chiral microstructures used in our experiments are comparable to OAM beams. Figure 1c is the sketch of optical vortices impinge on a chiral structure at a normal incidence by aligning to the central line. Compared to the linear energy flux of a plane wave, the Poynting vector of optical vortex follows a spiral trajectory twisting around the beam axis[26]. Hence, at a certain radial position of vortex beam, the angle between its Poynting vector and optical axis can be record as $\theta$ and $-\theta$ for right-helical wavefront (RHW) and left-helical wavefront (LHW), respectively. Furthermore, the chiral structure has a helical angle of $\alpha = \tan^{-1}[H/(\pi D)]$, where $H$ is its helical pitch and $D$ is its diametric position. Thus the interaction between helical wavefront and chiral structure can be simplified into a straightforward problem of the optical reflection of a slant plane wave incident on a sloped film (see insets in Fig. 1c). Considering that two slant plane waves with their incident angles of $\theta$-$\alpha$ and -$\theta$-$\alpha$ illuminate on the film, the difference of reflectance between two slant plane waves can be quantitatively evaluated as $g_R = 2 \times (R_{\theta-\alpha} - R_{-\theta-\alpha})/(R_{\theta-\alpha} + R_{-\theta-\alpha})$, where $R_{\theta-\alpha}$ and $R_{-\theta-\alpha}$ are the reflection coefficients of the two plane waves, respectively. For a homogenous and nonmagnetic thin film, the reflection coefficient of light with s-polarization is taken as a function of the incident angle[27]. For two mirror-symmetric sample with opposite sloped angle $\alpha$ and $-\alpha$, it can be proven (see Supplementary Information note 1) that $g_R(\theta, \alpha) = -g_R(\theta, -\alpha)$. Akin to sloped films illuminated by slant plane waves, the differences of reflectance of chiral structures with opposite helical angle are considerably symmetric under the illumination of optical vortices for realizing HD spectroscopy.

**Experimental set-up and reflection measurements.** The optical set-up for detecting HD in our configuration is depicted in Fig. 2a. A linearly polarized laser with a Gaussian mode at the wavelength of 800 nm is used in our experiments and then collimated to illuminate a fork hologram on a liquid-crystal spatial light modulator (SLM). In order to separate the first-order vortex beam, a tilted shift phase is added to the azimuthal phase hologram that has $l$ phase cycles around the axis (see Fig. S4). A 4$f$ system with an iris located at the confocal plane of $L_3$ and $L_4$ is employed to filter out undesired orders of diffraction. It should be noted that the sign of topological charge $l$ of vortex beam is reversed after three reflections upon mirrors[28].

The well-prepared vortex beam is then focused by a 100× objective lens to generate a doughnut-shape optical field. The chiral microstructures by analogy to spiral staircases are fabricated in polymer resist SZ2080 by direct laser writing (see Fig. S5). The chiral microstructures have a diameter $D$=17.4 μm, helical pitch $H$=21.4 μm, and thickness $t$=2.2 μm. An achiral cylindrical microstructure has also been fabricated as a control sample. Figures 2 b-d show the scanning electron microscope (SEM) images corresponding to left-handed, right-handed, and achiral microstructures after development. All the chiral and achiral microstructures are disc-shaped in the top-down view (see Fig. S6). The microstructure is aligned to the center of the focused vortex beam by using a high-precision nano-positioning stage. The reflected light from microstructure is collected by the same objective and finally recorded by a charge-coupled-device (CCD) camera (see Methods).

To avoid conventional CD effect, a linearly polarized light is used in our experiment. The interaction between helical wavefronts and chiral geometry can be obtained by measuring the asymmetric reflectance as a function of topological charge $l$. For the left-handed microstructure, light with RHW has stronger reflectance than that with LHW when the topological charge ranges $|l|$ from 5 to 65, as shown in Fig. 2e. In contrast, the right-handed microstructure obtains more reflectance under the LHW illumination compared with the RHW case (see Fig. 2f). Both results show that the reflectance from microstructure with different chirality behaves oppositely, which implies the emergence of HD phenomenon. As a control experiment, an achiral microstructure has the same reflectance for both cases of RHW and LHW illumination, as observed in Fig. 2g. Moreover, the difference of reflectance is independent on the polarization of vortex beam, and has been observed with both linearly and circularly polarized light (see Fig. S7).

**Measurement of HD spectrum.** It is instructive to compare qualitatively the CD response of two enantiomers from different systems with Kuhn's $g$-factor[29,30]. To evaluate the effect quantitatively, we calculate HD by the following dissymmetry factor defined as[18,24]

$$\text{HD}(\%) = \frac{I_R - I_L}{(I_R + I_L)/2} \times 100 \tag{1}$$

where $I_R$ and $I_L$ are the experimental reflection intensity under the RHW and LHW illumination, respectively. In Fig. 3, the measured HD spectra as a function of /l/ show positive, negative and zero values for left-handed, right-handed and achiral structures respectively, which indicates a clear discrimination between the two enantiomers. Left-handed microstructure produces a HD signal with a characteristic peak at $l \approx 32$ and is convergent to zero with $l > 65$ (see Supplementary Information table 1). Right-handed microstructure produces a vertically mirrored dip HD signal. The responses of both enantiomers are mirror images over the zero line, where the HD spectrum of achiral microstructure is located. Compared to the HD of ~0.6% range obtained by chiral molecules on unique plasmonic nanoparticle aggregates[18], the measured HD values of ~120% are significantly improved. Distinguishing from CD, the HD spectra of chiral microstructures are polarization-independent and significant as confirmed in our experiments with circular polarized light (see Figs. S8 and S9). Moreover, the conventional CD measurements require large amounts of analytes[31] or arrayed chiral metasurfaces[32] for enhancing the signal, whereas the HD response in our configuration is implemented directly with a single chiral structure. Therefore, the capability of detecting enantiomers with less quantity would be beneficial for limited production efficiency.

**Dependence of HD on the dimension of chiral structures.** In the HD phenomenon, the focused vortex beam is taken as a probe to detect the chirality of structures, so the focal behavior is important to understand the HD effects. Experimentally, the optical field distributions of tightly focused optical vortices with opposite topological charges $l = \pm 20$ at the focal plane are captured by the CCD, as shown in Fig. 4a. The optical profiles of focused vortex beams can be simulated by the vectorial Debye diffraction theory (see more details in Methods). As $E_z$ component is undetectable in an objective-based optical microscopy[33], we only consider the intensity and phase distribution predominately determined by $E_x$ component (see Fig. S10). The simulated optical vortices with opposite topological charge $l$ have the same optical field distribution and opposite azimuthal phase gradient, as illustrated in Fig. 4b. The diameter of ring profile $d$ is defined as the distance between two peaks on the cross-section. Both experimented and simulated results show the equal diameter $d$ for two focused vortex beams with opposite topological charge $l$. Figure 4c shows that the experimental diameters $d$ of focused vortex beams are linearly dependent on topological charge $|l|$, which is consistent with the predicated $|l|$ scaling law under tightly focused condition[34]. (See Fig. S11 and Supplementary Information note 2 for information on the linear function under the optical system.) It can also be observed in the simulated optical field cross-sections of optical vortices with varying positive and negative topological charge $l$, which are in excellent agreement with the experimental results, as shown in Figs. 4d and 4e.

The linearly variant diameters of focused vortex beams offer a path to unveil the connection between chiral geometry and HD spectra. With the increment of topological charge |*l*|, the HD spectrum at a given diameter *D* increases from zero to its maximum because the doughnut-shape profiles enlarges and their interacting area shifts from the center to outer region. But the HD spectrum begins to drop from the maximum when the outmost boundary of doughnut-shape profile is larger than *D*, which implies that one part of vortex beam is incident on the area outside the chiral microstructure. Intuitively, the reflected intensity profiles for vortex beams with opposite topological charges *l* provide a straightforward illustration for the HD generation (see Fig. S12).

To further elucidate the HD mechanism, we investigate the HD response under different diameter *D* and helical pitch *H* of chiral microstructures. Firstly, the shift of HD spectrum occurs over a change in diameter *D* of chiral microstructure. We experimentally measure the HD spectra of left-handed microstructures with fixed helical pitch *H*=21.4 μm and varying diameters *D* from 11.5 μm to 21.2 μm, as shown in Fig. 5a. As the diameters *D* of chiral microstructures increase, a shift in HD spectrum towards larger topological charge |*l*| is observed. When doughnut-shape profiles with large topological charge |*l*| exactly deviate from the chiral microstructure, the HD values approach to zero without interaction between optical vortices and structures, which are defined as critical points in HD spectra (see Figs. S13, S14). Due to the linearly variant diameters of focused vortex beams, the shift positions of critical points show a linear relationship with *D*, as shown in Fig. 5b. Although the experiment is implemented by using chiral microstructures with different *D*, their maximum HD almost keep the same value of ~120%. Secondly, the helical pitch *H* of chiral microstructures changes the helical angle *α* (according to its definition), which significantly influences the maximum HD values. Figure 5c depicts the HD spectra of left-handed microstructures with fixed diameter $D$=17.3 μm and varying helical pitch $H$ from 11.3 μm to 31.4 μm. In our measurements, the maximum HD is located at $H$=21.4 μm, where a strong interaction occurs by matching the scale of chiral microstructure and helical pitch of helical phase wavefront. However, the critical points maintain at $|l|$=60 because of the fixed diameter of chiral microstructures, as shown in Fig. 5d. The analyses of HD spectra can also be applied to guide the design of chiral structures and predict their HD properties qualitatively and quantitatively.

## Conclusions

In this study, we have observed a giant HD on a chiral meso-structure induced by optical vortex. The HD spectroscopy, based on the reflectance of vortex beams with topological charge $l$ from -75 to 75, is mirror-symmetric for two enantiomers, where the HD values are the same in magnitude but opposite in sign. The strong HD effect has a significant enhancement of 200 times compared with previous reports based on unique nanoparticle aggregates. The HD spectrum has tight dependence on the dimensions of chiral structures, such as diameter and helical pitch. Our results prove that the dichroism domain of optical properties on chiral structures can be expanded by the HD response for complementing traditional CD response. This technology offers an innovative method to detect the chirality of microscopic objects by using the OAM of light, and potentially benefits a wide spectroscopy of areas across chemistry, materials, biology and optics.

# Methods

**Optical apparatus.** The femtosecond laser source is a mode-locked Ti:sapphire ultrafast oscillator (Chameleon Vision-S, from Coherent Inc, USA) with a central wavelength of 800 nm, a pulse width of 75 fs, and a repetition rate of 80 MHz. The reflective liquid-crystal SLM (Pluto NIR-2, from Holoeye Photonics AG, Germany) has 1920 × 1080 pixels, with pixel pitch of 8 μm, on which CGHs with 256 gray levels can be displayed. The sample is mounted on a nanopositioning stage (E545, from Physik Instrumente (PI) GmbH & Co. KG, Germany) with nanometer resolution and a 200 μm×200 μm×200 μm moving range to precisely locate microstructures under optical microscopy.

**Geometry of chiral microstructures.** To define chiral microstructures using two-photon polymerization, we exploit the following parameterization in Cartesian coordinates:

$$\Omega(\varphi, R, \delta) = \begin{pmatrix} x(\varphi, R, \delta) \\ y(\varphi, R, \delta) \\ z(\varphi, R, \delta) \end{pmatrix} = \begin{pmatrix} \sqrt{R^2 + \delta^2} \times \cos[\pm\varphi + \tan^{-1}(\delta/R)] \\ \sqrt{R^2 + \delta^2} \times \sin[\pm\varphi + \tan^{-1}(\delta/R)] \\ H\varphi/2\pi \end{pmatrix} \quad \begin{matrix} \varphi \in [0, 2\pi] \\ R \in [0, D/2] \\ \delta \in [-\delta_0/2, \delta_0/2] \end{matrix} \quad (2)$$

which defines right-handed (+) and left-handed (-) microstructures of diameter $D$, cross-sectional width $\delta_0$=2 μm and helical pitch $H$.

**Sample preparation and characterization.** A commercially available zirconium-silicon hybrid sol-gel material (SZ2080), provided by IESL-FORTH (Greece), was used in our experiment and was negligibly shrinkable during structuring compared with other photoresists. The pre-baking process used to evaporate the solvent in the SZ2080 was set to a thermal platform at 100 ℃ for 45 min. After polymerization under illumination of a femtosecond laser, the sample was developed in 1-propanol for half an hour until the entire portion without polymerization was washed away. The SEM images were taken with a secondary electron scanning electron microscope (ZEISS EVO18) operated at an accelerating voltage of 10 keV after depositing ~10 nm gold.

**Details of the experiment measuring HD spectrum.** After positioning the chiral meso-structure to the center of optical vortex, the reflected intensity profiles by addressing their topological charges were caught by a CCD (Panasonic WV-BP334 camera having 768×576 pixels with the acquisition time of 500 ms). The laser power measured after an iris was 0.5 mW for clarity of the optical images. These optical images were gathered over 30 min per measurement of chiral meso-structure. This approach excludes time-dependent effects from the measured signals, showing the robust behavior.

**Numerical simulation of field distribution under the objective.** Under the paraxial approximation, the electric field of an optical vortex beam can be described as

$$\boldsymbol{E}_l^{in}(r,\varphi,z) = \sqrt{\frac{2}{\pi\omega^2(z)|l|!}} \left(\frac{\sqrt{2}r}{\omega(z)}\right)^{|l|} \exp\left(-\frac{r^2}{\omega^2(z)}\right) \exp(il\varphi) \boldsymbol{E}_x \qquad (3)$$

where $\omega(z)$ is the radius of beam, $\boldsymbol{E_x} = E_x \cdot \hat{\boldsymbol{x}}$ is the horizontal polarization vector, and $r, \varphi, z$ are the cylindrical coordinates[3]. When the beam waist (where $z = 0$ in Eq. (3)) is located at the back aperture plane of a high-NA objective lens, the electric field distribution on the focal plane can be simulated by using vectorial Debye diffraction theory[35,36]

$$\vec{\mathbf{E}}(r_2, \varphi_2) = \begin{bmatrix} E_x \cdot \hat{\boldsymbol{x}} \\ E_y \cdot \hat{\boldsymbol{y}} \\ E_z \cdot \hat{\boldsymbol{z}} \end{bmatrix}$$

$$= iC[\hat{\boldsymbol{x}} \quad \hat{\boldsymbol{y}} \quad \hat{\boldsymbol{z}}] \iint_\Omega \sin(\theta) E_i^{in}(\theta, \varphi) \sqrt{\cos\theta} \begin{bmatrix} 1 + (\cos\theta - 1)\cos^2\varphi \\ (\cos\theta - 1)\cos\varphi\sin\varphi \\ \sin\theta\cos\varphi \end{bmatrix} \exp\{ikr_2\sin\theta\cos(\varphi - \varphi_2)\}d\theta d\varphi \quad (4)$$

where $\vec{\mathbf{E}}(r_2, \varphi_2)$ is the electric field vector, $(r_2, \varphi_2)$ are the polar coordinates at the focal plane, $C$ is a constant, $\theta$ is the convergence angle of the objective, and $k$ is the wavenumber of light.

## Acknowledgements

This work was supported by the National Science Foundation of China (Nos. 61475149, 51675503, 61875181 and 61705085), the Fundamental Research Funds for the Central Universities (WK 2090090012, WK2480000002, WK2090090021 and WK2030380014), Youth Innovation Promotion Association CAS (2017495) and National Key R&D Program of China (2018YFB1105400). The authors thank Yaowei Huang and Jie Xu for useful stimulating discussion. K. H. thanks the support from CAS Poineer Hundred Talents Program.

## Author contributions

J.N. and D.W. conceived the idea and designed the experiments. Z.H., Y.H. and J.N. performed the simulations. J.N., Z.W. and Z.L. performed the measurements. Y.H., S.J., J.C. and J.L. analyzed the data. J.N., K.H., C.Q. and D.W. wrote the manuscript. All authors discussed the results and commented on the manuscript.

# Figures and captions

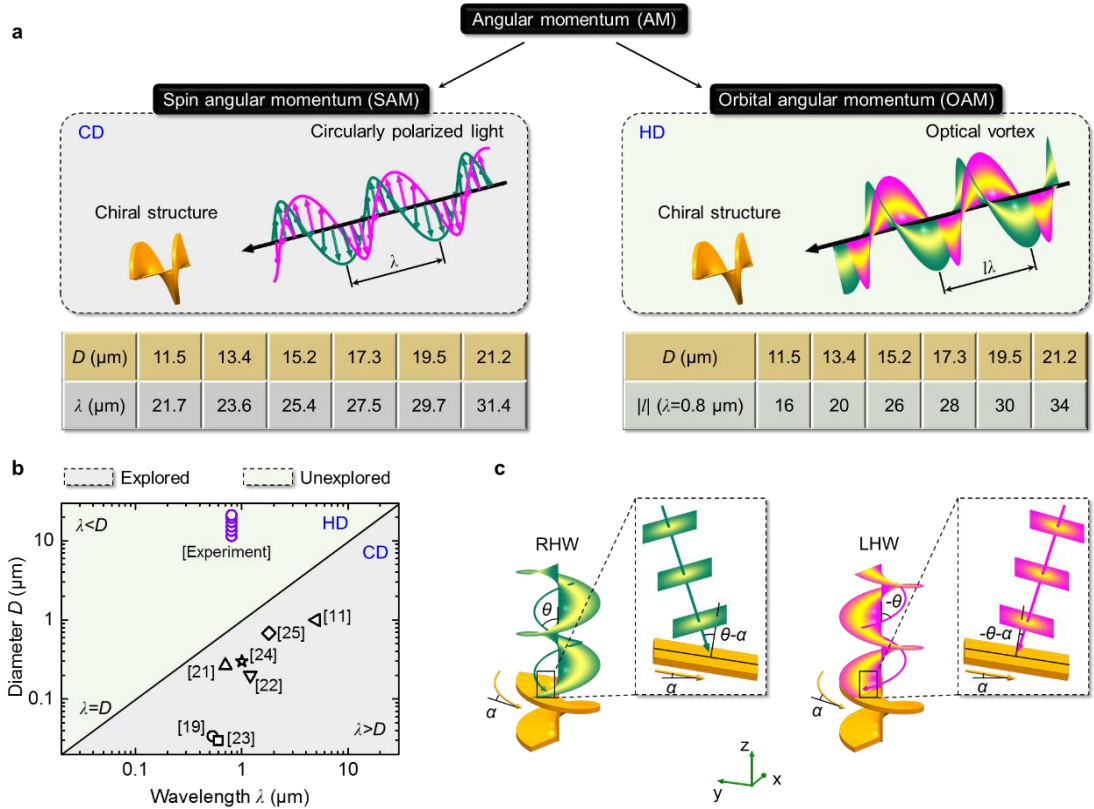

**Figure 1 | Principles of distinguishing chiral structures with OAM beam.** (**a**) Schematic plot of interaction between chiral structures and light with SAM and OAM. CD induced by circularly polarized light is normally at a wavelength $\lambda$ comparable or larger than the scale of chiral structure (left panel). The CD peak is red shifted by increasing the diameters $D$ of chiral microstructures (left table). HD induced by OAM beams locates at matching the scales of light beams and chiral structures (right panel). The HD peak is right shifted with larger chiral microstructures (right table). The wavelength used in our experiment is $\lambda$=800 nm. (**b**) Comparative overview of the CD and HD response area. Dark patterns represent the peaks of CD spectra found in previous literatures[11,19,21-25]; the black line indicates that wavelength $\lambda$ is equal to the

diameter of chiral structure *D*; the purple circles represent the results of HD obtained in this paper. (**c**) Illustration of OAM beams with RHW (left panel) and LHW (right panel) illuminating on a chiral structure. The Poynting vector (green and magenta arrows) screws around the beam axis. The helical angle of Poynting vector for RHW (LHW) and chiral structure are $\theta$ (-$\theta$), and $\alpha$, respectively. Insets illustrate two slant plane waves illuminating on a sloped sheet by unfolding the helical wavefront and chiral structure.

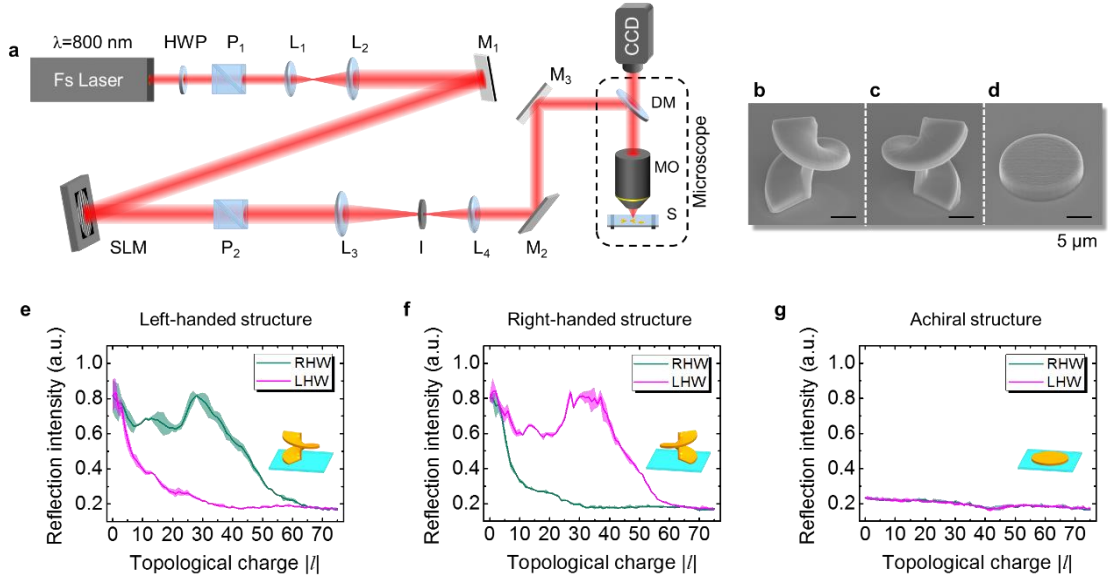

**Figure 2 | Experimental apparatus for generating OAM beams and detecting reflectance on chiral microstructures.** (**a**) A femtosecond laser is expanded by a telescope (lenses $L_1$-$L_2$) to match the size of an SLM, after its polarization and power adjusted by a half-wave plate (HWP) and a polarizer ($P_1$). The linearly polarized Gaussian light beam is modulated by the SLM after reflecting from a mirror ($M_1$). The desired vortex beam at the first-order diffraction is selected by an iris (I) and then introduced into a microscope system. After that, the vortex beam is focused on the sample (S) by a microscope objective (MO) with a high numerical aperture (NA=0.9). Finally, the reflected patterns are recorded by a charged-couple device (CCD) camera. DM, dichroic mirror. SEM images of left-handed (**b**), right-handed (**c**), and achiral (**d**) microstructures fabricated by direct laser writing in a polymer SZ2080. The corresponding measured reflectance on left-handed (**e**), right-handed (**f**) and achiral (**g**) microstructures as a function of topological charge $|l|$. Each solid line shows the mean value and the shading indicates their standard deviation of a set of three measurements with the same sample. a.u., arbitrary units.

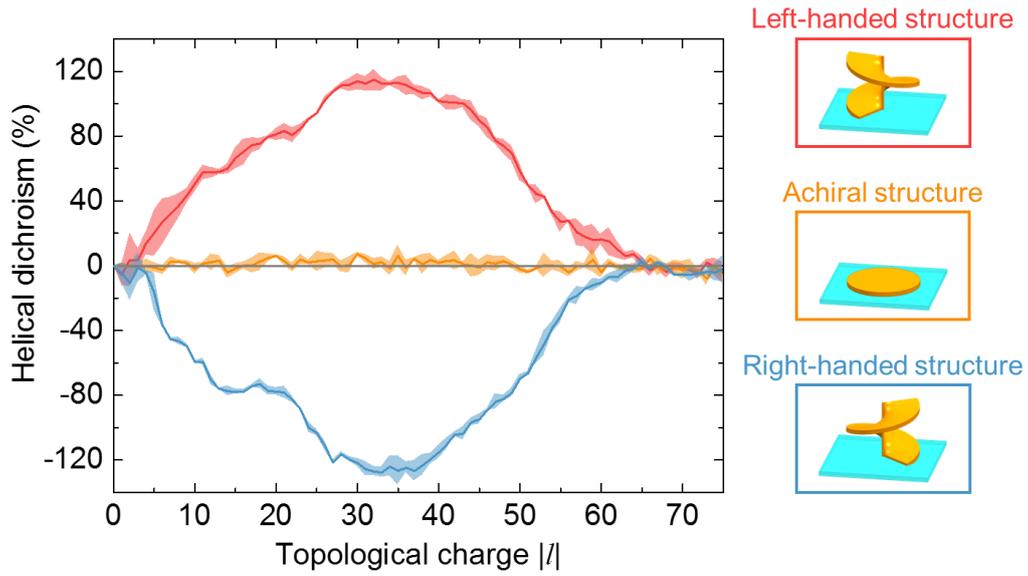

**Figure 3 | Helical dichroism measurements of chiral microstructures.** Symmetric distribution of HD between two chiral microstructures with opposite handedness. The grey straight line indicates HD=0 for guide-to-the-eye. The solid lines represent the mean values; shading indicates the standard deviation of a set of three measurements with the same sample.

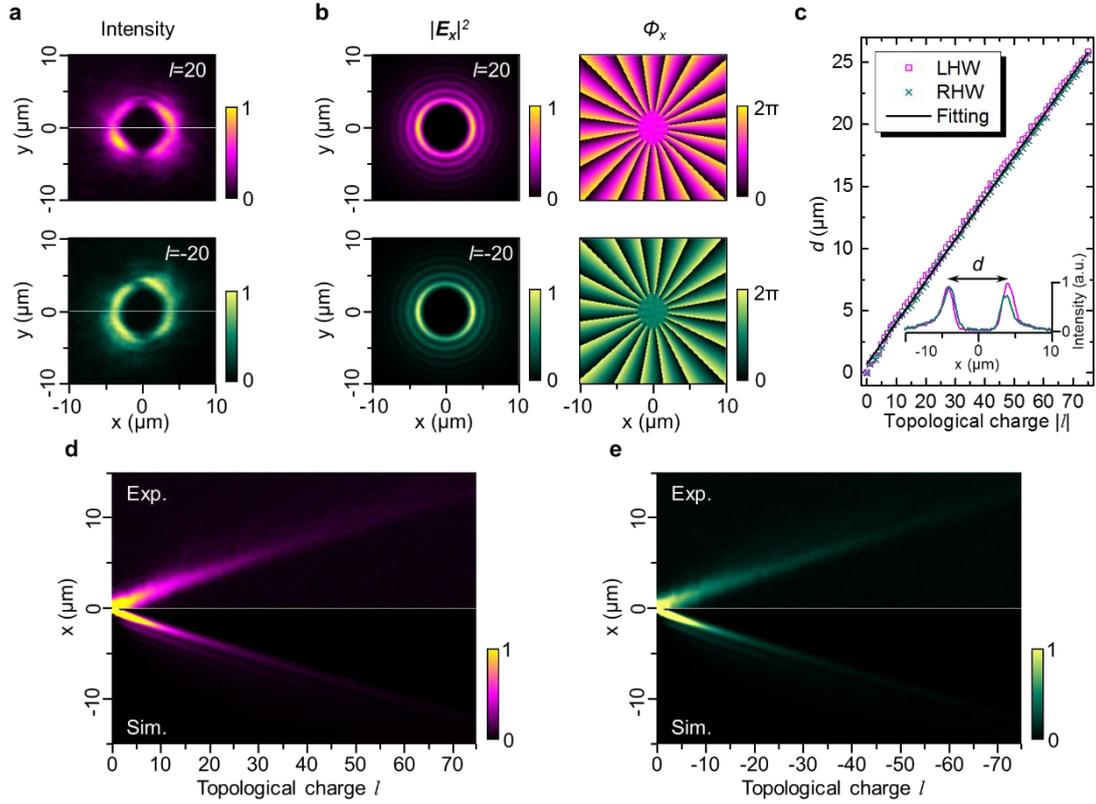

**Figure 4 | Experimental observation and simulation of focused OAM beams.** (**a**) Experimental measurement of intensity profiles on the focal plane with OAM value $l$=20 (top panels) and -20 (bottom panels). (**b**) Simulated intensity and phase distributions of optical vortices corresponding to (a). (**c**) Measured diameter $d$ of optical vortices as a function of topological charge $|l|$. The solid line represents the linear fitting of measured diameter $d$ (black line) proportional to $|l|$. Inset, the cross-section of measured intensity profiles in (a). (**d-e**) 2D mapping cross-sectional intensity profiles of optical vortices with varying topological charge $l$ from -75 to 75 in experiment and simulation.

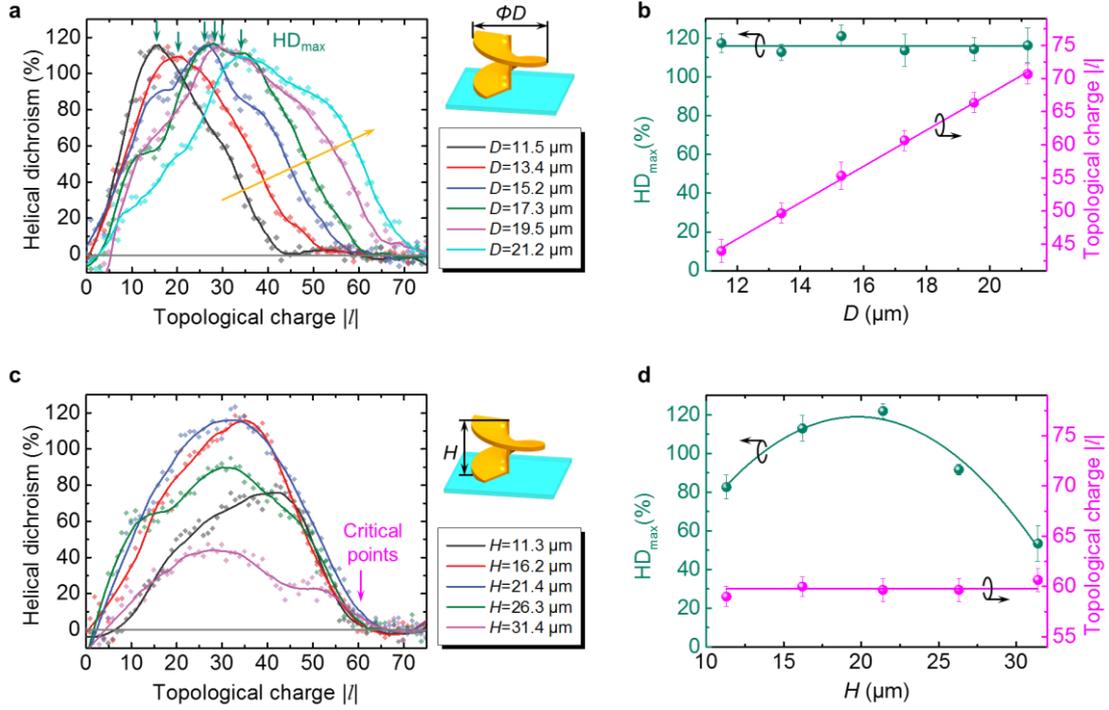

**Figure 5 | Structure-property relationships for HD spectra.** (**a**) Measured HD spectra of chiral microstructures with fixed helical pitch $H$=21.4 μm and different diameters $D$ from 11.5 to 21.2 μm. The yellow arrow indicates increasing diameters $D$ and green arrows indicate the maximum HD values. The HD spectra are right shifted by increasing the diameters $D$ of chiral microstructures. (**b**) Measured maximum HD values (green dots) and topological charge $|l|$ of critical points (magenta dots, defined in c) as a function of diameters $D$. (**c**) Measured HD spectra of chiral microstructures with fixed diameter $D$=17.3 μm and different helical pitch $H$ from 11.3 to 31.4 μm. Critical points indicate that HD values exactly return to zero. (**d**) Measured maximum HD values (green dots) and topological charge $|l|$ of critical points (magenta dots) as a function of helical pitch $H$. The curves are drawn as a guide to the eye. Error bars are the standard deviation of three measurements with the same sample.

24